\documentclass[prd,aps,preprint,draft,showpacs]{revtex4}
\usepackage{epsf}
\usepackage[dvips]{color}
\usepackage{dcolumn}
\newcommand{\Tr}{\mbox{Tr}}
\newcommand{\VEV}[1]{\left\langle #1\right\rangle}

\newcommand{\ie}{{\it i.e.\ }}

\newcommand{\SXPT}{S$\chi$PT}
\newcommand{\rSXPT}{rS$\chi$PT}
\newcommand{\diag}{\rm diag}

\begin{document}
\title{Scalar Meson Spectroscopy with Lattice Staggered Fermions}
\author{Claude Bernard}
\affiliation{
Physics Department, Washington University, St.~Louis, MO 63130, USA
}
\author{Carleton DeTar and Ziwen Fu}
\affiliation{
Physics Department, University of Utah, Salt Lake City, UT 84112, USA
}
\author{Sasa Prelovsek }
\affiliation{
Department of Physics, University of Ljubljana, Jadranska 19, Ljubljana, Slovenia \\
and J.~Stefan Institute, Jamova 39, Ljubljana, Slovenia
}
\date{\today}

\begin{abstract}
With sufficiently light up and down quarks the isovector ($a_0$) and
isosinglet ($f_0$) scalar meson propagators are dominated at large
distance by two-meson states.  In the staggered fermion formulation of
lattice quantum chromodynamics, taste-symmetry breaking causes a
proliferation of two-meson states that further complicates the
analysis of these channels.  Many of them are unphysical artifacts of
the lattice approximation.  They are expected to disappear in the
continuum limit.  The staggered-fermion fourth-root procedure has its
purported counterpart in rooted staggered chiral perturbation theory
(\rSXPT).  Fortunately, the rooted theory provides a strict framework
that permits the analysis of scalar meson correlators in terms of only
a small number of low energy couplings.  Thus the analysis of the
point-to-point scalar meson correlators in this context gives a useful
consistency check of the fourth-root procedure and its proposed chiral
realization.  Through numerical simulation we have measured
correlators for both the $a_0$ and $f_0$ channels in the ``Asqtad''
improved staggered fermion formulation in a lattice ensemble with
lattice spacing $a = 0.12$ fm.  We analyze those correlators in the
context of \rSXPT\ and obtain values of the low energy chiral
couplings that are reasonably consistent with previous determinations.
\end{abstract}
\pacs{11.15.Ha, 12.38.Gc, 12.39.Fe, 14.40.Cs}

\maketitle

\section{Introduction}

The recent evident successes of numerical simulations of QCD with
improved staggered fermions demand a thorough examination of its most
controversial ingredient, namely, using fractional powers of the
determinant to simulate the correct number of quark species (the
``fourth root trick'').  The procedure is known to introduce
nonlocalities and violations of unitarity at nonzero lattice spacing
\cite{Bernard:2006ee}.  If these problems do not vanish in the
continuum limit, they may even place the theory in an unphysical
universality class.  There are, however, strong theoretical arguments
\cite{Shamir:2006nj,Bernard:2006zw,Sharpe:2006re,Bernard:2006qt} that
the fourth-root trick is valid, \ie that it produces QCD in the
continuum limit.

One may also test the fourth-root procedure numerically.
One can, for example, check that taste symmetry gets restored 
as the lattice spacing gets smaller, by looking at the eigenvalue
spectrum \cite{Durr:2003xs,Follana:2004sz,Durr:2004as,Durr:2004ta,Follana:2005km}, the Dirac operator \cite{Bernard:2005gf},
or the pion spectrum \cite{Bernard:2006wx}. Alternatively,
low-energy results of
staggered fermion QCD simulations can be compared
with predictions of rooted staggered
chiral perturbation theory (\rSXPT) \cite{Aubin:2003mg,Aubin:2003uc}.
Since staggered chiral perturbation theory becomes standard chiral perturbation theory
in the continuum
limit, agreement between rooted QCD and (\rSXPT) at nonzero lattice
spacing would suggest that, at least for low energy or long-range
phenomena, lattice artifacts produced by the fourth root approximation
are as harmless as those produced by partial quenching.  Partial
quenching also induces unitarity violations, but they disappear
in the limit of equal valence and sea quark masses.

There are two recent tests of agreement between rooted
staggered fermion QCD and \rSXPT: (1) Measurements of the light
pseudoscalar meson masses and decay constants in partially quenched
and full staggered fermion QCD fit well to expressions derived from
\rSXPT\ \cite{Aubin:2004fs}.  A byproduct of this fit is a
determination of the low energy couplings of the theory.  (2) The
topological susceptibility measured in full QCD agrees reasonably well
with predictions of \rSXPT\ \cite{Billeter:2004wx}.

In the present work we examine scalar meson correlators in full QCD
and compare their two-meson content with predictions of \rSXPT\@.
Since the appearance of the two-meson intermediate state is a 
consequence of the fermion determinant, an analysis of this correlator
provides a direct test of the fourth root recipe.  The $a_0$ channel has been
studied recently in staggered fermion QCD by the MILC collaboration
\cite{Aubin:2004wf} and UKQCD collaboration \cite{Gregory:2005yr}.
Both groups found that the correlator appeared to contain states with
energies well below possible combinations of physical mesons.

A simple explanation of the nonstandard features of the scalar correlators is provied
by \rSXPT\ 
\cite{Prelovsek:2005qc,Prelovsek:2005rf}.  In that theory all
pseudoscalar mesons come in multiplets of 16 tastes.  The pattern of
mass splittings is predicted by the theory.  The $\pi$ and $K$
multiplets are split in similar ways.  The $\eta$ and $\eta^\prime$
mesons, however, are peculiar, because their masses are shifted by the
axial $U(1)$ anomaly.  Since the anomaly is a taste singlet, only the
taste singlet $\eta$ and $\eta^\prime$ acquire approximately physical
masses.  Some of the remaining members of the $\eta$ multiplet remain
degenerate with the pions.  According to taste symmetry selection
rules, any two mesons coupling to a taste-singlet $a_0$ must have the
same taste.  But all tastes are equally allowed.  Among other states,
the taste singlet $a_0$ couples to the Goldstone pion (pseudoscalar
taste) and an $\eta$, also with pseudoscalar taste and of the same
mass.  This spurious two-body state at twice the mass of the Goldstone
boson accounts for the anomalous low-energy component in that channel.

This explanation raises concerns.  Clearly, only the taste singlet
$\eta$ approximates the physical state, since it is the only member of
the multiplet subject to the anomaly.  So if the other $\eta$'s are
not allowed as external states, we have violated unitarity in the
sense that some intermediate states are not allowed as external
states.  Further examination of the taste multiplets in the
intermediate states reveals that in addition to the several unphysical
$\pi \eta$ taste combinations, there is a negative norm ``ghost''
contribution in the taste singlet $\eta$ meson leg
\cite{Bardeen:2001jm}.  Remarkably, all lattice artifacts resolve
themselves in the continuum limit, however.  The taste multiplets
become degenerate, the two-body states merge, and the ghost state
cancels the spurious taste combinations, leaving only the
taste-singlet mesons.  To achieve this cancellation requires following
the rules of flavor counting in \rSXPT\@.

In the present work we extend the analysis of
Ref.~\cite{Prelovsek:2005qc,Prelovsek:2005rf} and carry out a
quantitative comparison of measured correlators and predictions of
\rSXPT\@.  Despite the considerable complexity of channels with
dozens of spectral components, the chiral theory models the
correlators precisely in terms of only a small number of low energy
couplings, which we may determine through fits to the data.

This article is organized as follows.  Following a review of some
needed results from \SXPT\ in Sec \ref{sec:sxpt}, we derive the chiral
predictions for the $a_0$ and $f_0$ in Sec \ref{sec:correlators}.  We
present results of our fits to the predicted forms in Sec
\ref{sec:results} and conclude in Sec \ref{sec:conclude}.


\section{Elements of staggered chiral perturbation theory}
\label{sec:sxpt}

In this section we give a brief review of rooted staggered chiral
perturbation theory with particular emphasis on the tree-level
pseudoscalar mass spectrum.  We obtain the rooted version of the
theory through the replica trick, according to which each quark
flavor, $u$, $d$, and $s$, comes in four tastes and is repeated $n_r$
times \cite{Aubin:2003rg}.  We calculate various quantities in the
replicated theory, and in the final step, we set $n_r = 1/4$ to obtain
the correct flavor counting.

The low energy effective chiral theory is formulated in terms of the
meson field
\begin{equation}
  \Phi = \sum_{b=1}^{16} \frac{1}{2}T^b \phi^b
\end{equation}
where $T^b = \{1, \xi_5,i\xi_5\xi_\mu \ldots{}\}$ are Dirac gamma
matrices and $\phi^b$ is a $3n_r \times 3n_r$ matrix with rows and
columns labeled by the flavor and replica index $ur$, $dr$, and $sr$.
The staggered chiral action is written in terms of the unitary matrix
$\Sigma = \exp(2i\Phi/f)$:
\begin{equation}
  S(\Sigma,m) = \int d^4y \left\{
  \frac{f^2}{8}\Tr(\partial_\mu \Sigma^\dagger \partial^\mu \Sigma)
  - \frac{\mu f^2}{4}\Tr({\cal M}\Sigma^\dagger + {\cal M}^\dagger \Sigma)
   + {m_0^2 \over 2}\phi_{0I}^2
   + a^2 {\cal V}(\Sigma) \right\}~.
\label{eq:SXPT}
\end{equation}
The low energy couplings at this order are $f$, $\mu$, and the quark
mass matrix ${\cal M} = I_t \otimes I_r \diag(m_u, m_d, m_s)$, where
$I_t$ is the unit matrix in taste space and $I_r$ is the unit matrix
in replica space.  The axial anomaly appears through the mass term
$m_0^2$.  It involves the flavor-singlet taste-singlet field
$\phi_{0I} = \sum_{f,r} \phi^I_{fr,fr}/\sqrt{3n_r}.$ The
taste-breaking term ${\cal V}$ is a linear combination of operators
\cite{Lee:1999zx,Aubin:2003mg,Aubin:2003uc}
\begin{equation}
  -{\cal V}(\Sigma) = \sum C_i {\cal O}_i,
\end{equation}
where
\begin{eqnarray}
  {\cal O}_1 &=& \Tr\left( T_{0,5} \Sigma T_{0,5} \Sigma^\dagger \right) \\
  {\cal O}_{2V} &=& {1 \over 4} \left[
      \Tr( T_{0,\mu} \Sigma ) \Tr( T_{0,\mu} \Sigma ) + \mbox{h.c.}\right] \\
  {\cal O}_{2A} &=& {1 \over 4} \left[
      \Tr( T_{0,\mu 5} \Sigma ) \Tr( T_{0,5\mu} \Sigma ) + \mbox{h.c.}\right] \\
  {\cal O}_3 &=& {1 \over 2}\left[
      \Tr( T_{0,\mu} \Sigma T_{0,\mu} \Sigma ) + \mbox{h.c.}\right] \\
  {\cal O}_4 &=& {1 \over 2}\left[
      \Tr( T_{0,\mu 5} \Sigma T_{0,5\mu} \Sigma ) + \mbox{h.c.}\right] \\
  {\cal O}_{5V} &=& {1 \over 2} \left[
      \Tr( T_{0,\mu} \Sigma ) \Tr( T_{0,\mu} \Sigma^\dagger )\right] \\
  {\cal O}_{5A} &=& {1 \over 2} \left[
      \Tr( T_{0,\mu 5} \Sigma ) \Tr( T_{0,5 \mu} \Sigma^\dagger )\right] \\
  {\cal O}_6 &=& \sum_{\mu < \nu}
       \Tr\left( T_{0,\mu\nu} \Sigma T_{0,\nu\mu} \Sigma^\dagger \right)  \ \ .
\end{eqnarray}

Without the anomaly and taste-breaking term the tree-level masses of
the pseudoscalar mesons with quark flavor content $f,f^\prime$ are, as
usual,
\begin{equation}
  M^2_{f,f^\prime,b} = \mu (m_f + m_{f^\prime}).
\end{equation}

The taste-breaking term splits the nonisosinglet states ($\pi_b$ and
$K_b$) to give
\begin{equation}
    M_{f,f^\prime,b}^2 = \mu (m_f + m_{f^\prime}) + a^2 \Delta_b,
\end{equation}
where to leading order the multiplets split five ways,
\begin{eqnarray}
  \Delta_5 &=& 0 \nonumber \\
  \Delta_{\mu 5} &=& \frac{16}{f^2}(C_1 + 3 C_3 +  C_4 + 3 C_6) \\
  \Delta_{\mu \nu} &=& \frac{16}{f^2}(2 C_3 + 2 C_4 + 4 C_6) \nonumber \\
  \Delta_\mu &=& \frac{16}{f^2}(C_1 + C_3 + 3 C_4 + 3 C_6) \nonumber \\
  \Delta_I &=& \frac{16}{f^2}(4 C_3 + 4 C_4), \nonumber
\end{eqnarray}
which we label $P$, $A$, $T$, $V$, and $I$, respectively.  This
predicted multiplet pattern has been well confirmed in simulations
\cite{Bernard:2001av,Aubin:2004wf}.

We will be working with degenerate $u$ and $d$ quarks ($m_u = m_d =
m_\ell$), so it will be convenient to introduce the notation
\begin{eqnarray}
  M^2_{Ub} &=& 2\mu m_\ell  + a^2 \Delta_b \nonumber \\
  M^2_{Sb} &=& 2\mu m_s  + a^2 \Delta_b  \\
  M^2_{Kb} &=& \mu(m_\ell +  m_s)  + a^2 \Delta_b \nonumber
\end{eqnarray}

The isosinglet states ($\eta$ and $\eta^\prime$) are modified both by
the taste-singlet anomaly and by the two-trace (quark-line hairpin)
taste-vector and taste-axial-vector operators ${\cal O}_{2V}$, ${\cal
O}_{2A}$, ${\cal O}_{5V}$ and ${\cal O}_{5A}$.  When $m_0^2$ is large,
in the taste-singlet sector we obtain the usual result
\begin{eqnarray}
   M_{\eta, I}^2 &=& \frac{1}{3}M_{UI}^2 + \frac{2}{3}M_{SI}^2 \nonumber \\
   M_{\eta^\prime, I} &=&  {\cal O}(m_0^2).
\end{eqnarray}
In the taste-axial-vector sector we have
\begin{eqnarray}
M^2_{\eta A}&=&\frac{1}{2}[M^2_{UA}+M^2_{SA}+3n_r\delta_A-Z_A] \nonumber \\
M^2_{\eta^\prime A}&=&\frac{1}{2}[M^2_{UA}+M^2_{SA}+3n_r\delta_A+Z_A] \\
Z^2_A&=&(M^2_{SA}-M^2_{UA})^2-2 n_r \delta_A (M^2_{SA}-M^2_{UA})+9 n_r^2\delta^2_A, \nonumber
\end{eqnarray}
where $\delta_A = a^2 \delta^\prime_A = a^2 16(C_{2A} - C_{5A})/f^2$,
and likewise for $A \rightarrow V$.

In the taste-pseudoscalar and taste-tensor sectors, in which is there is no mixing
of the isosinglet states, the $\eta_b$ and $\eta'_b$ by definition 
have quark content $(\bar uu+\bar dd)/\sqrt{2}$
and $\bar s s$, respectively, and masses
\begin{equation}
 M_{\eta, b}^2 = M_{Ub}^2 \ ;\qquad  M_{\eta', b}^2 = M_{Sb}^2
\end{equation}

In Table \ref{tab:pseudoscalar} we list the masses of the resulting taste multiplets
for the lattice ensemble used in the present study with taste-breaking
parameters $\delta_A$ and $\delta_V$ determined in
Ref.~\cite{Aubin:2004wf,Aubin:2004fs}.

\begin{table}[h]
\begin{center}
\begin{tabular}{lllll}
  $b$ & $\pi_b$ & $K_b$ & $\eta_b$ & $\eta^\prime_b$ \\
 \hline
  P   & 0.1594 & 0.3652 & 0.1594 & 0.4927 \\ 
  A   & 0.2342 & 0.4036 & 0.1843 & 0.5129 \\
  T   & 0.2694 & 0.4250 & 0.2694 & 0.5384 \\
  V   & 0.2966 & 0.4428 & 0.2825 & 0.5491 \\
  I   & 0.3205 & 0.4592 & 0.4958 & $-$   
\end{tabular}
\end{center}
\caption{Masses of pseudoscalar meson taste multiplets in lattice
units for the MILC coarse ($a = 0.12$ fm) lattice ensemble $\beta =
6.76$, $am_{ud} = 0.005$, $am_s = 0.05$, as measured or inferred from
measured masses and splittings. The mass of the $\eta^\prime_I$
depends on the anomaly parameter $m_0$. \label{tab:pseudoscalar}}
\end{table}


\section{Scalar correlators from \SXPT}
\label{sec:correlators}

In this section we rederive the ``bubble'' contribution to the $a_0$
channel of Ref.~\cite{Prelovsek:2005rf}, using the language of the
replica trick \citep{Damgaard:2000gh,Aubin:2003mg}, and then extend the
result to the $f_0$ channel.

We match the point-to-point scalar correlators in chiral low energy
effective theory and staggered fermion QCD by matching the Green's
functions, which are defined through the generating functionals of the
respective theories:
\begin{equation}
\frac{\partial^2 \log Z}
       {\partial m_{f,f^\prime}(y) 
         \partial m_{e^\prime,e}(0)}~.
\label{eq:matching}
\end{equation}
For this purpose the quark mass term $\diag(m_u,m_d,m_s)$ is converted
to a local meson source $m_{ff^\prime}(y)$ (including flavor
off-diagonal terms) in both \SXPT\ and QCD.


\subsection{Scalar correlator in staggered fermion QCD}

First we review the construction of the needed correlators in
staggered lattice QCD\@, where the generating functional is
\begin{equation}
  Z(m_{ff^\prime}) = \int dU \exp[-S_g(U)]\det[M(U,m_{ff^\prime})]^{1/4}~.
\label{eq:QCDgen}
\end{equation}
Here $U$ are the gauge link variables, $S_g(U)$ is the gauge action,
and $M$ is the fermion matrix including flavor components.
We work on a lattice of spacing $a$ and dimension $L^3\times N_t$ and
label sites by the integer four-vector $x_\mu$.  Hypercubes of size
$2^4$ are similarly labeled by $y_\mu$, so $x_\mu = 2y_\mu +
\eta_\mu$.

Staggered fermion meson correlators can be defined in the
one-component basis of the Grassman color vector field $\chi_f(x)$ or
in the spin-taste basis of the field $q_f^{a\alpha}(y)$ with spin label
$\alpha$ and taste label $a$.  The fields are related through
\begin{eqnarray}
   q_f^{a\alpha}(y) &=&
    \frac{1}{8}\sum_\eta \Gamma^{a\alpha}_\eta \chi_f(2y+\eta) \nonumber \\
   \chi_f(2y+\eta) &=& 2 \Tr [\Gamma^\dagger_\eta q_f(y)]
\end{eqnarray}
where $\Gamma_\eta =
\gamma_0^{\eta_0}\gamma_1^{\eta_1}\gamma_2^{\eta_2}\gamma_3^{\eta_3}$,
and the sum over $\eta$ runs over sites in the $2^4$ hypercube labeled by
$y$.  The lattice $y$ has spacing $A = 2a$.

For constructing the meson correlators via the functional derivative 
(\ref{eq:matching}) we need to introduce the source term into Lagrangian
\begin{equation}
  S_m = a^4 \sum_x \bar \chi_f(x) \chi_{f^\prime}(x) m_{f,f^\prime}(x)~.
\end{equation}
To express the source in term of the spin-taste basis we use the relation
\begin{equation}
  a^4 \bar \chi_f(2y+\eta)\chi_{f^\prime}(2y+\eta) = 
     \frac{A^4}{16} \sum_\Gamma \zeta(\Gamma,\eta) \rho_{f,f^\prime,\Gamma}(y)
\label{eq:relation1}
\end{equation}
with
\begin{equation}
  \zeta(\Gamma,\eta) = \Tr(\Gamma_\eta^\dagger \Gamma^\dagger
       \Gamma_\eta \Gamma)/4
\end{equation}
and
\begin{equation}
\rho_{f,f^\prime,\Gamma}(y) = 
    \bar q_f(y)\Gamma\otimes \Gamma^* q_{f^\prime}(y)~.
\end{equation}
The direct product $\Gamma_S \otimes \Gamma^*_T$ acts on
spin and taste components, respectively.
So we obtain 
\begin{equation}
S_m =  A^4 \sum_{y,\Gamma} 
\rho_{f,f^\prime,\Gamma}(y) m_{f,f^\prime,\Gamma}(y)
\end{equation}
with
\begin{equation}
  m_{f,f^\prime,\Gamma}(y) = \frac{1}{16} \sum_\eta \zeta(\Gamma,\eta) 
      m_{f,f^\prime}(2y+\eta)
\end{equation}

The desired source for the scalar density, $m_{f,f^\prime, I}(y)$, has
$\Gamma = I$ and $\zeta(I,\eta) = 1$. It is the component of
$m_{f,f^\prime}(x)$ that is constant over a $2^4$ hypercube.  The
other terms $m_{f,f^\prime, \Gamma}(y)$ are sources for the other
local staggered mesons.

A particular correlator is obtained by differentiating the generating
functional with respect to the appropriate source mass terms.  The
general two-point function is, then,
\begin{equation}
     \left.\frac{\partial^2 \log Z}
       {\partial m_{f,f^\prime,\Gamma}(y) 
         \partial m_{e^\prime,e,\Gamma^\prime}(0)}
    \right|_{m_{ff^\prime}(x) = \delta_{ff^\prime}m_f} = A^8\VEV{\bar \rho_{f,f^\prime,\Gamma}(y)\rho_{e^\prime,e,\Gamma^\prime}(0)}.
\label{eq:scalarcorr}
\end{equation}
The above quantity  will be calculated for $\Gamma=I$ also within \SXPT\ 
below.

Now, we need to relate the quantity (\ref{eq:scalarcorr}) to the
correlator generated from the code.  In practice the simulated
correlator is computed from a point source at the origin
\begin{equation}
  O_{e,e^\prime,\rm src} = a^3 \bar \chi_{e^\prime}(0) \chi_e(0) 
    = \frac{A^3}{8} \sum_\Gamma \rho_{e,e^\prime\Gamma}(0),
\label{eq:src}
\end{equation}
and a single time-slice sink operator at time $\tau = 2t + \eta_0$,
\begin{equation}
   O_{f,f^\prime \rm sink}(\vec y,\tau) 
   = a^3 \sum_{\vec \eta} \bar \chi_{f^\prime}(2\vec y+\vec \eta,\tau) 
    \chi_{f}(2\vec y+\vec \eta, \tau) 
   = A^3 [\rho_{f,f^\prime,I}(2 \vec y,t) + 
   (-)^{\eta_0} \rho_{f,f^\prime,05}(2 \vec y, t)],
\label{eq:sink}
\end{equation}
where we have used relation (\ref{eq:relation1}), 
$\zeta(I,\eta) = 1$ and $\zeta(05,\eta) =
(-)^{\eta_0}$.  Note that the sink operator is defined on spatial
cubes $\vec y$ but all time slices $\tau$.

In this language the computed correlator is
\begin{equation}
   C_{f,f^\prime;e,e^\prime}(\vec p,\tau a) = 
    \sum_{\vec y} \exp(i \vec p \cdot \vec y A)
       \VEV{\bar O_{f,f^\prime\rm sink}(\vec y,\tau)
        O_{e,e^\prime\rm src}}.
\label{eq:defstaggcorr}
\end{equation}
The meson taste is conserved, so the correlator separates into
nonoscillating and oscillating components for a taste-singlet scalar
contribution and a taste-axial-vector pseudoscalar meson contribution,
respectively.
\begin{equation}
  C_{f,f^\prime;e,e^\prime}(\vec p,\tau a) = 
   C_{f,f^\prime;e,e^\prime,I}(\vec p,\tau a) + 
     (-)^\tau C_{f,f^\prime;e,e^\prime,05}(\vec p,\tau a),
\end{equation}
where
\begin{equation}
 C_{f,f^\prime;e,e^\prime,\Gamma}(\vec p,\tau a) = \frac{A^6}{8}
    \sum_{\vec y} \exp(i \vec p \cdot \vec y A)
 \VEV{\rho_{f,f^\prime,\Gamma}(2 \vec y,t)\rho_{e,e^\prime,\Gamma}(0)}.
\label{eq:corrvsdens}
\end{equation}

The correlator has a quark-line-connected part and may also have a
quark-line disconnected part:
\begin{equation}
   C_{f,f^\prime;e,e^\prime,\Gamma}(\vec p,\tau a) = 
    C_{f,f^\prime;e,e^\prime,\Gamma,\rm conn}(\vec p,\tau a) + 
    C_{f,f^\prime;e,e^\prime,\Gamma,\rm disc}(\vec p,\tau a)
\end{equation}
The quark-line disconnected part appears only in the taste-singlet
isosinglet correlator.

To compute the correlator we need to express it in terms of quark
propagators.  So we start from the definition of the correlator in
Eq.~(\ref{eq:defstaggcorr}), substitute the definitions of the
operators in Eqs.~(\ref{eq:src}) and (\ref{eq:sink}) and use the
relation
\begin{eqnarray}
a^8\VEV{\bar \chi_f(2y+\eta)\chi_{f^\prime}(2y+\eta) 
   \bar \chi_e(\eta^\prime)\chi_{e^\prime}(\eta^\prime)}
  &=& \left.\frac{\partial^2 \log Z}
     {\partial m_{f,f^\prime}(2y+\eta)
         \partial m_{e^\prime,e}(\eta^\prime)}
    \right|_{m_{ff^\prime}(x) = \delta_{ff^\prime}m_f} 
 \label{eq:onecorrvsST} \\
  &=& \frac{A^8}{256}\sum_{\Gamma,\Gamma^\prime}
  \zeta(\Gamma,\eta) \zeta(\Gamma^\prime,\eta^\prime)
\VEV{\bar \rho_{f,f^\prime,\Gamma}(y)\rho_{e^\prime,e,\Gamma^\prime}(0)}~,
 \nonumber
\end{eqnarray} 
which follows from identity 
\begin{equation}
  \frac{\partial}{\partial m(2y + \eta)} = \frac{1}{16} \sum_\Gamma
   \zeta(\Gamma,\eta)\frac{\partial}{\partial m_\Gamma(y)}~.
\end{equation} 
Finally we arrive at the point-to-point correlators
\begin{eqnarray}
 C_{f,f^\prime;e,e^\prime,\rm conn}(\vec p,\tau a) &=& 
    -\sum_{\vec x} (-)^x \exp(i \vec p \cdot \vec x a)
    \VEV{\Tr[M_f^{-1}(\vec x,\tau;0,0)
             M_{f^\prime}^{-1^\dagger}(\vec x,\tau;0,0)]}
    \delta_{ef}\delta_{e^\prime f^\prime} \nonumber \\
 C_{f,f^\prime;e,e^\prime,\rm disc}(\vec p,\tau a)  &=& 
   \frac{1}{4}
    \sum_{\vec x} \exp(i \vec p \cdot \vec x a)
      \VEV{\Tr[M_f^{-1}(\vec x,\tau;\vec x,\tau)]
           \Tr[M_e^{-1^\dagger}(0,0;0,0)]}
    \delta_{ee^\prime}\delta_{ff^\prime}, \nonumber
\end{eqnarray}
where we have used Eq.~(\ref{eq:QCDgen}) and 
the normalization $M = 2D + 2am$ for the Dirac
matrix.  We keep the momentum $p$ small, so we can neglect variation
of the exponential over the hypercube.

As it is computed, at zero momentum the quark-line disconnected
correlator includes the vacuum disconnected piece:
\begin{equation}
   C_{f,e,0} = \frac{L^3}{4}\VEV{\Tr[M_{f}^{-1}(0,0;0,0)]}
     \VEV{\Tr[M_{e}^{-1}(0,0;0,0)]}.
\label{eq:vacdisc}
\end{equation}


\subsection{Scalar correlator in \SXPT}

The continuum generating functional for scalar correlators in \SXPT
\begin{equation}
Z_{\rm SXPT}(m_{ff^\prime}) = \int [d\Sigma] \exp[-S(\Sigma,m_{ff^\prime})],
\end{equation}
where $S(\Sigma,m_{ff^\prime})$ is given by Eq.~(\ref{eq:SXPT}).  We
do not include explicit scalar meson fields in the chiral Lagrangian,
but add their contributions in the final expressions.
To match the functional derivatives (\ref{eq:matching}) 
we approximate the continuum
integration in the chiral theory with a sum over hypercubic volumes of
size $A^4$ and differentiate with respect to a constant source inside
that volume.  In this case $m_{f,f^\prime,I}(y)=m_{f,f^\prime}(y)$.  
The source is also constant over replicas of the same
flavor.  The space-time volume equals that of QCD, namely, $A^4
(L/2)^3 N_t/2$ for $A = 2a$. We use the integer four vector $y$ to
label the hypercubes in the chiral theory. The functional derivative 
in \SXPT\  is 
\begin{equation}
\frac{\partial^2 \log Z_{\rm SXPT}}
       {\partial m_{f,f^\prime}(y) 
         \partial m_{e^\prime,e}(0)}=A^8\mu^2\sum_{r,r^\prime}
   \VEV{\Tr_t\left(\Phi^2(y)\right)_{fr,f^\prime r}
   \Tr_t\left(\Phi^2(0)\right)_{er^\prime ,e^\prime r^\prime}}~.
\label{eq:matching1}
\end{equation}
At tree level the action (\ref{eq:SXPT}) has no explicit
quark-antiquark scalar meson fields, but it generates the
two-pseudoscalar-meson ``bubble'' terms in the correlator.  The
functional derivative (\ref{eq:matching1}) corresponds to
(\ref{eq:scalarcorr}) with $\Gamma=I$. We use $B$ to denote the bubble
contribution corresponding to the correlator (\ref{eq:corrvsdens})
\begin{eqnarray}
   B_{f,f^\prime;e,e^\prime,I}(\vec p, tA) &=& 
   \frac{A^6}{8}\sum_{\vec y} \exp( i\vec p \cdot \vec y A)
   \VEV{\rho_{f,f^\prime,I}(y)\rho_{e,e^\prime,I}(0)} \\
   &=&
  \frac{A^6}{8}\sum_{\vec y} \exp(i \vec p \cdot \vec y A)
   \mu^2\sum_{r,r^\prime}
   \VEV{\Tr_t\left(\Phi^2(y)\right)_{fr,f^\prime r}
   \Tr_t\left(\Phi^2(0)\right)_{er^\prime ,e^\prime r^\prime}}.
  \nonumber
\end{eqnarray}
We introduce its time Fourier transform
\begin{equation}
   B_{f,f^\prime;e,e^\prime,I}(p) = \sum_{t=0}^{N_t/2} \exp(ip_0 t A)
   B_{f,f^\prime;e,e^\prime,I}(\vec p,tA).
\end{equation}

At tree level the vacuum expectation value reduces through Wick
contractions to products of meson two-point functions.  
In momentum space we have, generically, the Euclidean correlator
\begin{equation}
  \VEV{\phi(y)\phi(0)} = \frac{1}{A^4 (L/2)^3(N_t/2)}
   \sum_k \exp(ik \cdot y A) \VEV{\phi(-k)\phi(k)}
\end{equation}
where $\VEV{\phi(-k)\phi(k)} = 1/(k^2 + m^2)$.  So
\begin{equation}
  \VEV{\phi(y)\phi(0)} = 
    \frac{1}{A^3 (L/2)^3}\sum_{\vec k}
    \frac{\exp[-E(\vec k)tA + i\vec k \cdot \vec y A]}{2 E(\vec k)}
\end{equation}
for $E(\vec k) = \sqrt{|\vec k|^2 + m^2}$.  In terms of momentum
components, the general term in the correlator becomes
\begin{eqnarray}
B_{f,f^\prime;e,e^\prime,I}(p) 
&=&
  \frac{A^6 \mu^2}{8}\sum_{y} \exp(i p \cdot y A) \sum_{g,s,r,b} 
   \sum_{g^\prime,s^\prime,r^\prime,b^\prime}
   \VEV{\phi^b_{fr,gs}(\vec y, t)\phi^b_{gs,f^\prime r}(\vec y, t)
   \phi^{b^\prime}_{e r^\prime, g^\prime s^\prime}(0)
    \phi^{b^\prime}_{g^\prime s^\prime e^\prime r^\prime}(0)} \nonumber \\
&=& \frac{\mu^2}{8 (L/2)^3 (N_t/2)A^2}
  \sum_k \sum_{g,s,r,b}
   \sum_{g^\prime,s^\prime,r^\prime,b^\prime}\left[
    \VEV{\phi^b_{fr,gs}(-k)
     \phi^{b^\prime}_{e r^\prime, g^\prime s^\prime}(k)} \right. \nonumber \\
&&
   \VEV{\phi^b_{gs,f^\prime r}(k-p)
     \phi^{b^\prime}_{g^\prime s^\prime e^\prime r^\prime}(p-k)}
    \nonumber \\
 &+& \left.
    \VEV{\phi^b_{fr, gs}(-k)
    \phi^{b^\prime}_{g^\prime s^\prime e^\prime r^\prime}(k)} 
   \VEV{\phi^b_{gs,f^\prime r}(k - p)
        \phi^{b^\prime}_{e r^\prime, g^\prime s^\prime}(p-k)}\right].
\end{eqnarray}
We have used the fact that the bubble term, by definition, does not include
the vacuum disconnected piece corresponding to Eq.~(\ref{eq:vacdisc}).

There are two types of two-point functions, namely, the connected
two-point function for all tastes:
\begin{equation}
\VEV{\phi^b_{gs,fr}(-k)
   \phi^b_{f^\prime r^\prime,g^\prime s^\prime}(k)}_{\rm conn} =
 \frac{\delta_{r,r^\prime}\delta_{f,f^\prime}
       \delta_{g,g^\prime}\delta_{s,s^\prime}}{k^2 + M^2_{fg,b}},
\label{eq:conn}
\end{equation}
and the additional disconnected contribution for the taste-singlet,
taste-axial-vector, and taste-vector mesons.  For the taste singlet it
is
\begin{equation}
 \VEV{\phi^I_{gs,fr}(-k)
   \phi^I_{f^\prime r^\prime,g^\prime s^\prime}(k)}_{\rm disc} =
  -\frac{\delta_{r,s}\delta_{r^\prime,s^\prime}
   \delta_{f,g}\delta_{f^\prime,g^\prime}}{3n_r} \frac{k^2 + M^2_{SI}}
    {(k^2 + M^2_{UI})(k^2 + M^2_{\eta I}) }.
\label{eq:disc}
\end{equation}
Here we have already decoupled the taste-singlet $\eta'$ by taking $m_0^2\to \infty$.
The disconnected contribution for the taste-axial-vector meson is
\begin{equation}
 \VEV{\phi^A_{gs,fr}(-k)
   \phi^A_{f^\prime r^\prime,g^\prime s^\prime}(k)}_{\rm disc} =
  -\delta_{r,s}\delta_{r^\prime,s^\prime}
   \delta_{f,g}\delta_{f^\prime,g^\prime} 
    \frac{\delta_A(k^2 + M^2_{SA})}
    {(k^2 + M^2_{UA})(k^2 + M^2_{\eta A}) (k^2 + M^2_{\eta^\prime A})}~,
\label{eq:discA}
\end{equation}
and there is a similar contribution for the taste-vector meson.

It is convenient to carry out a partial fraction expansion of the
disconnected contributions as follows:
\begin{equation}
 \VEV{\phi^I_{gs,fr}(-k)
   \phi^I_{f^\prime r^\prime,g^\prime s^\prime}(k)}_{\rm disc} =
  -\frac{\delta_{r,s}\delta_{r^\prime,s^\prime}
   \delta_{f,g}\delta_{f^\prime,g^\prime}
  }{3n_r}
   \left(
     \frac{3/2}{k^2 + M^2_{UI} } -
           \frac{1/2}{k^2 + M^2_{\eta I}}  \right)
\label{eq:disc-expand}
\end{equation}
and
\begin{equation}
 \VEV{\phi^A_{gs,fr}(-k)
   \phi^A_{f^\prime r^\prime,g^\prime s^\prime}(k)}_{\rm disc} =
  -\delta_{r,s}\delta_{r^\prime,s^\prime}
   \delta_{f,g}\delta_{f^\prime,g^\prime} \delta_A
   \left( \frac{g_U}
    {k^2 + M^2_{UA}} + 
    \frac{g_\eta}
    {k^2 + M^2_{\eta A}} + 
    \frac{g_{\eta^\prime}}
    {k^2 + M^2_{\eta^\prime A}}\right)~,
\end{equation}
where
\begin{eqnarray}
 g_U & = &  \frac{M^2_{SA} - M^2_{UA}}
   {(M^2_{\eta A} - M^2_{UA})(M^2_{\eta^\prime A} - M^2_{UA})} \nonumber \\
 g_\eta & = &  \frac{M^2_{SA} - M^2_{\eta A}}
          {(M^2_{UA} - M^2_{\eta A})(M^2_{\eta^\prime A} - M^2_{\eta A})} \\
 g_{\eta^\prime} & = & \frac{M^2_{SA} - M^2_{\eta^\prime A}}
   {(M^2_{UA} - M^2_{\eta^\prime A})(M^2_{\eta A} - M^2_{\eta^\prime A})} \nonumber
\end{eqnarray}
In the language of Refs.~\cite{Aubin:2003mg,Aubin:2003uc}, $g_U$,
$g_\eta$, and $g_{\eta^\prime}$ are simply the residues for
Eq.~(\ref{eq:discA}). Similarly, the factors of $3/2$ and $-1/2$ in
Eq.~(\ref{eq:disc-expand}) are the residues for Eq.~(\ref{eq:disc}).

\subsection{Isovector $a_0$ correlator}

We now specialize to the isovector $a_0$ correlator.  We consider, for
simplicity, the $u\bar d$ flavor state.  Only the quark-line-connected
contribution appears in the QCD correlator
\begin{equation}
 B_{a_0}(\vec p, \tau a) =  B_{u,d;d,u}(\vec p,\tau a) = 
    -\sum_{\vec x}(-)^x \exp(i\vec p \cdot \vec x)
      \VEV{\Tr[M_u^{-1}(\vec x,\tau;0,0)
             M_d^{-1^\dagger}(\vec x,\tau;0,0)]}.
\end{equation}
In terms of the meson fields, the bubble correlator is
(for $\tau a = t A$)
\begin{eqnarray}
B_{u,d;u,d,I}(\vec p,t A) &=& \frac{A^6 \mu^2}{8}
  \sum_{\vec y} \exp(i\vec p \cdot \vec y A)
  \sum_{r,s,f,b}
  \sum_{r^\prime,s^\prime,f^\prime,b^\prime} \\
  &&
  \VEV{\phi^b_{ur,fs}(\vec y,t)\phi^b_{fs,dr}(\vec y,t) 
   \phi^{b^\prime}_{dr^\prime,f^\prime s^\prime}(0)
   \phi^{b^\prime}_{f^\prime s^\prime,ur^\prime}(0)}
  \nonumber
\label{eq:fourmeson}
\end{eqnarray}
After carrying out the Wick contractions and switching to momentum
space we get
\begin{eqnarray}
  B_{a_0}(p) &=& \frac{\mu^2}{8 A^2(L/2)^3 (N_t/2)}\left\{ n_r^2
  \sum_{f,b}\sum_k\left[
  \frac{1}{k^2 + M^2_{fu,b}} 
  \frac{1}{(k+p)^2 + M^2_{fu,b}}
  \right] \right.
  \nonumber \\ &-& 
  4 n_r \sum_k \left[
  \frac{1}{(k+p)^2 + M^2_{U I}}
  \frac{1}{3n_r} \frac{k^2 + M^2_{SI}}
    {(k^2 + M^2_{UI})(k^2 + M^2_{\eta I}) }
  \right] \nonumber \\ &-& 
  4 n_r \sum_k \left[
  \frac{4\delta_A}{(k+p)^2 + M^2_{U A}}
    \frac{k^2 + M^2_{SA}}
    {(k^2 + M^2_{UA})(k^2 + M^2_{\eta A}) (k^2 + M^2_{\eta^\prime A})}
  \right] \nonumber \\ &-& \left.
  4 n_r \sum_k \left[
  \frac{4\delta_V}{(k+p)^2 + M^2_{U V}}
    \frac{k^2 + M^2_{SV}}
    {(k^2 + M^2_{UV})(k^2 + M^2_{\eta V}) (k^2 + M^2_{\eta^\prime V})}
  \right]\right\}.
\label{eq:a0bubblemom}
\end{eqnarray}
Notice, in particular, the negative weight threshold in the second
term and the spurious taste-nonsinglet $\pi \eta$ thresholds involving
Goldstone-boson-like members of the $\eta$ taste multiplet.

In the continuum limit, in which taste-symmetry is restored, we have
\begin{eqnarray}
  B_{a_0}(p) &=& \frac{\mu^2}{8 A^2(L/2)^3 (N_t/2)}\left\{ 16 n_r^2
  \sum_{f}\sum_k\left[
  \frac{1}{k^2 + M^2_{fu}} 
  \frac{1}{(k+p)^2 + M^2_{fu}}
  \right] \right.
  \nonumber \\&-& \left.
  \frac{4}{3}\sum_k \left[
  \frac{1}{(k+p)^2 + M^2_{U}}
   \frac{k^2 + M^2_{S}}
    {(k^2 + M^2_{U})(k^2 + M^2_{\eta}) }
  \right] \right\}.\nonumber
\end{eqnarray}
Here the total contribution from pairs of light states with mass $M_U$ is
proportional to
\begin{eqnarray}
  (32 n_r^2 - 2) \,,
\end{eqnarray}
which vanishes when $n_r = 1/4$.  The negative norm threshold has
neatly canceled the unphysical thresholds.  The surviving thresholds
are the physical $\bar K K$ and taste singlet $\pi \eta$.

\subsection{Isosinglet $f_0$ correlator}

In this case we use the isosinglet operator $(\rho_{uu,I} +
\rho_{dd,I})/\sqrt{2}$.  We have both quark-line-connected and
quark-line-disconnected contributions
\begin{eqnarray}
 B_{f_0}(\vec p, \tau a) &=& B_{f_0,\rm conn}(\vec p, \tau a) 
   + B_{f_0,\rm disc}(\vec p, \tau a)  \\
 B_{f_0,\rm conn}(\vec p, \tau a) &=& \frac{1}{2}
 [B_{u,u;u,u,\rm conn}(\vec p,\tau a) + B_{d,d;d,d,\rm conn}(\vec p,\tau a)] 
   \nonumber \\ &=& 
    -\sum_{\vec x} (-)^x \exp(i \vec p \cdot \vec x a)
    \VEV{\Tr[M_u^{-1}(\vec x,\tau;0,0)
             M_u^{-1^\dagger}(\vec x,\tau;0,0)]} \\
 B_{f_0,\rm disc}(\vec p,\tau a)  &=& \frac{1}{2}
 [B_{u,u;u,u,\rm disc}(\vec p,\tau a) + B_{u,u;d,d,\rm disc}(\vec p,\tau a) 
   \nonumber \\ &+&
 B_{d,d;u,u,\rm disc}(\vec p,\tau a) + B_{d,d;d,d,\rm disc}(\vec p,\tau a)]
   \nonumber\\  &=&  \frac{1}{2}
    \sum_{\vec x} \exp(i \vec p \cdot \vec x a)
      \VEV{\Tr[M_u^{-1}(\vec x,\tau;\vec x,\tau)]
           \Tr[M_u^{-1^\dagger}(0,0;0,0)]}~. 
\end{eqnarray}
The weight of the disconnected part is $n_f/4$ for $n_f = 2$
degenerate flavors for the state.  The connected part of the correlator
is identical to the full $a_0$ correlator.

In terms of the meson fields, the bubble correlator is
(for $\tau a = t A$)
\begin{eqnarray}
B_{f_0}(\vec p, t A)
 &=& \frac{ \mu^2 A^6}{8} 
  \sum_{\vec y} \exp(i\vec p \cdot \vec y A)
  \sum_{r,s,f,b}
  \sum_{r^\prime,s^\prime,f^\prime,b^\prime} \nonumber \\
  && \frac{1}{2}\left[
  \VEV{\phi^b_{ur,fs}(t)\phi^b_{fs,ur}(t) 
   \phi^{b^\prime}_{ur^\prime,f^\prime s^\prime}(0)
   \phi^{b^\prime}_{f^\prime s^\prime,ur^\prime}(0)}
 \right. \nonumber \\ &+& 
  \VEV{\phi^b_{ur,fs}(t)\phi^b_{fs,ur}(t)
   \phi^{b^\prime}_{dr^\prime,f^\prime s^\prime}(0)
   \phi^{b^\prime}_{f^\prime s^\prime,dr^\prime}(0)}
 +  \VEV{\phi^b_{dr,fs}(t)\phi^b_{fs,dr}(t)
   \phi^{b^\prime}_{ur^\prime,f^\prime s^\prime}(0)
   \phi^{b^\prime}_{f^\prime s^\prime,ur^\prime}(0)}
  \nonumber \\ &+& \left.
  \VEV{\phi^b_{dr,fs}(t)\phi^b_{fs,dr}(t)
   \phi^{b^\prime}_{dr^\prime,f^\prime s^\prime}(0)
   \phi^{b^\prime}_{f^\prime s^\prime,dr^\prime}(0)}
  \right]
\label{eq:fourmesonf0}
\end{eqnarray}
In momentum space the correlator becomes
\begin{eqnarray}
  B_{f_0}(p) &=& \frac{\mu^2}{8 A^2(L/2)^3 (N_t/2)}\left\{ n_r^2
  \sum_{f,b}\sum_k\left[
  \frac{1}{k^2 + M^2_{fu,b}} 
  \frac{1}{(k+p)^2 + M^2_{fu,b}}
  \right] \right.
  \nonumber \\ &+& 2 n_r^2
  \sum_b\sum_k\left[
  \frac{1}{k^2 + M^2_{U b}} 
  \frac{1}{(k+p)^2 + M^2_{U b}}
  \right]\\ &-& 
  4n_r\sum_k \left[
  \frac{1}{(k+p)^2 + M^2_{U I}}
  \frac{1}{3n_r} \frac{k^2 + M^2_{SI}}
    {(k^2 + M^2_{UI})(k^2 + M^2_{\eta I}) }
  \right] \nonumber \\ &-& 
  4n_r \sum_k \left[
  \frac{4\delta_A}{(k+p)^2 + M^2_{U A}}
    \frac{k^2 + M^2_{SA}}
    {(k^2 + M^2_{UA})(k^2 + M^2_{\eta A}) (k^2 + M^2_{\eta^\prime A})}
  \right] \nonumber \\ &-& 
  4n_r \sum_k \left[
  \frac{4\delta_V}{(k+p)^2 + M^2_{U V}}
    \frac{k^2 + M^2_{SV}}
    {(k^2 + M^2_{UV})(k^2 + M^2_{\eta V}) (k^2 + M^2_{\eta^\prime V})}
  \right] \nonumber \\ &+& 
  4 n_r^2  \sum_k  \left[
  \frac{1}{3n_r} \frac{(k+p)^2 + M^2_{SI}}
    {[(k+p)^2 + M^2_{UI}][(k+p)^2 + M^2_{\eta I}] }
  \frac{1}{3n_r} \frac{k^2 + M^2_{SI}}
    {(k^2 + M^2_{UI})(k^2 + M^2_{\eta I}) }
  \right] \nonumber \\ &+& 
  4 n_r^2 \sum_k  \left[
  \frac{4 \delta_A[(k+p)^2 + M^2_{SA}]}
{[(k+p)^2 + M^2_{UA}][(k+p)^2 + M^2_{\eta A}] 
   [(k+p)^2 + M^2_{\eta^\prime A}]}
   \right. \nonumber \\ &\times & \left.
    \frac{\delta_A(k^2 + M^2_{SA})}
    {(k^2 + M^2_{UA})(k^2 + M^2_{\eta A}) (k^2 + M^2_{\eta^\prime A})}
  \right] \nonumber \\ &+& 
  4 n_r^2 \sum_k  \left[
  \frac{4 \delta_V[(k+p)^2 + M^2_{SV}]}
{[(k+p)^2 + M^2_{UV}][(k+p)^2 + M^2_{\eta V}] 
  [(k+p)^2 + M^2_{\eta^\prime V}]}
   \right. \nonumber \\ &\times & \left.\left.
    \frac{\delta_V(k^2 + M^2_{SV})}
    {(k^2 + M^2_{UV})(k^2 + M^2_{\eta V}) (k^2 + M^2_{\eta^\prime V})}
  \right]\right\}  . \nonumber
\label{eq:f0bubblemom}
\end{eqnarray}
In terms of valence quark world lines the first five terms are
quark-line connected and the last three are disconnected.

In the continuum limit we have
\begin{eqnarray}
  B_{f_0}(p) &=& \frac{\mu^2}{8 A^2(L/2)^3 (N_t/2)}\left\{ 16 n_r^2
  \sum_{f}\sum_k\left[
  \frac{1}{k^2 + M^2_{fu}} 
  \frac{1}{(k+p)^2 + M^2_{fu}}
  \right] \right.
  \nonumber \\ &+& 32 n_r^2 \sum_k \left[
  \frac{1}{k^2 + M^2_{U}} 
  \frac{1}{(k+p)^2 + M^2_{U}}
  \right] \\ &-& 
  4 n_r \sum_k \left[
  \frac{1}{(k+p)^2 + M^2_{U}}
  \frac{1}{3n_r} \frac{k^2 + M^2_{S}}
    {(k^2 + M^2_{U})(k^2 + M^2_{\eta}) }
  \right] \nonumber \\ &+& 
  \left. 4 n_r^2  \left[
  \frac{1}{3n_r} \frac{(k+p)^2 + M^2_{S}}
    {[(k+p)^2 + M^2_{U}][(k+p)^2 + M^2_{\eta}] }
  \frac{1}{3n_r} \frac{k^2 + M^2_{S}}
    {(k^2 + M^2_{U})(k^2 + M^2_{\eta}) }
  \right]\right\} . \nonumber
\end{eqnarray}
The two-pion threshold $(p+k)^2 + M_U^2 = 0$ and $k^2 + M_U^2 = 0$
has a weight proportional to
\begin{eqnarray}
  (64 n_r^2 - 1)\mu^2.
\end{eqnarray}
When $n_r = 1/4$ the weight is $3$ (for three physical pion channels).
Thus, once again, only physical thresholds survive the continuum limit.

\subsection{Single-flavor staggered fermions}

Single-flavor QCD has no Goldstone bosons.  The low-lying pseudoscalar
(call it the $\eta^\prime$) is lifted by the anomaly.  With the
staggered fermion action, however, only the taste-singlet
$\eta^\prime$ is lifted by the anomaly.  The other 15 members of the
taste multiplet (call them $\eta$) remain light.  The member with
pseudoscalar taste is an exact Goldstone boson.  Such a spectrum would
seem to spell trouble for the rooted theory.  It is interesting to
examine the scalar meson ($f_0$) correlator to see how the
corresponding rooted chiral theory heals itself in the continuum
limit.

Call the single replicated flavor $u$.  The connected meson correlator
is as before [Eq.~(\ref{eq:conn})].  We choose not to decouple the
taste-singlet $\eta'$ in this case because it is the only physical
meson.  The disconnected correlator for the taste singlet is then
\begin{equation}
 \VEV{\phi^I_{gs,fr}(-k)
   \phi^I_{f^\prime r^\prime,g^\prime s^\prime}(k)}_{\rm disc} =
  \frac{\delta_{r,s}\delta_{r^\prime,s^\prime}
   \delta_{f,g}\delta_{f^\prime,g^\prime}}{n_r}\left[ - \frac{1}
    {(k^2 + M^2_{UI})} + \frac{1}{k^2 + M^2_{\eta^\prime I}} \right].
\end{equation}
Similarly, the disconnected correlators for the taste axial vector and
taste vector can be written as
\begin{equation}
 \VEV{\phi^A_{gs,fr}(-k)
   \phi^A_{f^\prime r^\prime,g^\prime s^\prime}(k)}_{\rm disc} =
  \delta_{r,s}\delta_{r^\prime,s^\prime}
   \delta_{f,g}\delta_{f^\prime,g^\prime}\frac{-\delta_A}
    {(k^2 + M^2_{UA})(k^2 + M^2_{\eta^\prime A})}~.
\end{equation}
and ($A \rightarrow V$), where in this case $M^2_{\eta^\prime A} = 
M^2_{UA} + n_r\delta_A$, and similarly for $M^2_{\eta^\prime V}$.

With these changes the $f_0$ correlator becomes
\begin{eqnarray}
  B_{f_0}(p) &=& \frac{\mu^2}{8A^2 (L/2)^3 (N_t/2)}\left\{ 2 n_r^2
  \sum_{b}\sum_k\left[
  \frac{1}{k^2 + M^2_{U b}} 
  \frac{1}{(k+p)^2 + M^2_{U b}}
  \right] \right.
  \nonumber \\ &-&
  4n_r\sum_k 
  \frac{1}{(k+p)^2 + M^2_{U I}}
  \frac{1}{n_r} \left[\frac{1}
    {k^2 + M^2_{UI}} - \frac{1}{k^2 + M^2_{\eta^\prime I}}
  \right] \nonumber \\ &+& 
  4n_r \sum_k \left[
  \frac{\delta_A}{[(k+p)^2 + M^2_{U A}]
    (k^2 + M^2_{UA})(k^2 + M^2_{\eta^\prime A})}
  \right] \nonumber \\ &+& 
  4n_r \sum_k \left[
  \frac{\delta_V}{[(k+p)^2 + M^2_{U V}]
    (k^2 + M^2_{UV})(k^2 + M^2_{\eta^\prime V})}
  \right]  \\ &+& 
  2 n_r^2  \sum_k  
  \frac{1}{n_r} \left[ \frac{1}
    {(k+p)^2 + M^2_{UI}} - \frac{1}{(k+p)^2 + M^2_{\eta^\prime I}} \right]
  \frac{1}{n_r}\left[ \frac{1}
    {k^2 + M^2_{UI}} - \frac{1}{k^2 + M^2_{\eta^\prime I}}
  \right] \nonumber \\ &+& 
  2 n_r^2 \sum_k  \left[
  \frac{\delta^2_A}
{[(k+p)^2 + M^2_{UA}]
   [(k+p)^2 + M^2_{\eta^\prime A}]
   (k^2 + M^2_{UA})(k^2 + M^2_{\eta^\prime A})}
  \right] \nonumber \\ &+& \left.
  2 n_r^2 \sum_k  \left[
  \frac{\delta^2_V}
{[(k+p)^2 + M^2_{UV}][(k+p)^2 + M^2_{\eta^\prime V}]
   (k^2 + M^2_{UV})(k^2 + M^2_{\eta^\prime V})}
  \right]\right\}  . \nonumber
\end{eqnarray}
We note that a simplified version of our result (setting the discretization corrections from
$\delta_A$ and $\delta_V$ to zero) was presented previously \cite{Bernard:2006zw}.
In the continuum limit the would-be-Goldstone thresholds become
degenerate with the negative-norm threshold, with a net weight
proportional to
%
%
%
\begin{equation}
  (32 n_r^2  - 2) \ .
\end{equation}
When $n_r = 1/4$, the would-be Goldstone bosons decouple from the
$f_0$ correlator, leaving only the physical high-lying $\eta^\prime \eta^\prime$
channel.


\section{Simulations and Results}
\label{sec:results}

In this work we analyzed the 0.12 fm ensemble of 510 $24^3 \times 64$
gauge configurations generated in the presence of $2+1$ flavors of
Asqtad improved staggered quarks with bare quark masses $am_{ud} =
0.005$ and $am_s = 0.05$ and bare gauge coupling $10/g^2 = 6.76$
\cite{Aubin:2004wf}.  

We set valence quark masses equal to the sea quark masses.  Table
\ref{tab:pseudoscalar} gives the pseudoscalar masses used in our fits with the
exception of the masses $\eta_A$, $\eta^\prime_A$, $\eta_V$,
$\eta^\prime_V$.  Those masses vary with the fit parameters $\delta_A$ and
$\delta_V$.

For the light quark Dirac operator $M_u$, we measured the point-to-point
quark-line connected correlator
\begin{equation}
   C_{\rm conn}(\vec p,\tau) = 
    \sum_{\vec x}(-)^x \cos(\vec p \cdot \vec x)
      \VEV{\Tr[M_u^{-1}(\vec x,\tau;0,0)M_u^{-1^\dagger}(\vec x,\tau;0,0)]}
\end{equation}
and point-to-point quark-line disconnected correlator
\begin{equation}
   C_{\rm disc}(\vec p,\tau) = 
    \sum_{\vec x}(-)^x \cos(\vec p \cdot \vec x)
      \VEV{\Tr M_u^{-1}(\vec x,\tau;\vec x,\tau)
             \Tr M_u^{-1}(0,0;0,0)}.
\end{equation}
In the latter case we use noisy estimators based on random $Z(2)$
color vectors \cite{Dong:1993pk} $\eta_k$ for $k = 1,\ldots{}N =
200$:
\begin{eqnarray}
  \Tr M_u^{-1}(\vec x,\tau;\vec x,\tau) \Tr M_u^{-1}(0,0;0,0)
    &\approx & \frac{1}{N(N-1)}
    \sum_{k \ne k^\prime, y, y^\prime} \bar \eta_k(\vec x,\tau) 
      M_u^{-1}(\vec x,\tau;y) \eta_k(y) \\
     &\times&
       \bar \eta_{k^\prime}(0, 0) 
       M_u^{-1}(0,0; y^\prime)\eta_{k^\prime}(y^\prime) \nonumber.
\end{eqnarray}
In terms of these correlators the $a_0$ and $f_0$ correlators are
\begin{eqnarray}
  C_{a_0}(\vec p,\tau) &=& C_{\rm conn}(\vec p,\tau) \nonumber\\
  C_{f_0}(\vec p,\tau) &=& C_{\rm conn}(\vec p,\tau) - 
   \frac{1}{2}C_{\rm disc}(\vec p,\tau)
\end{eqnarray}

Correlators in each channel were measured at five momenta $\vec p =
(0,0,0)$, $(1,0,0)$, $(1,1,0)$, $(1,1,1)$, and $(2,0,0)$.  All ten
correlators were then fit to the following model
\begin{eqnarray}
C_{a_0}(\vec p,\tau) &=& C_{{\rm meson},a_0}(\vec p,\tau)  + B_{a_0}(\vec p,\tau) 
  \nonumber \\
C_{f_0}(\vec p,\tau) &=& C_{{\rm meson},f_0}(\vec p,\tau)  + B_{f_0}(\vec p,\tau)
\label{eq:fitfcn}
\end{eqnarray}
where
\begin{eqnarray}
  C_{{\rm meson},a_0}(\vec p,\tau) &=& b_{a_0}(p)\exp[-E_{a_0}(p) \tau] + 
   b_{\pi,A}(p)(-)^\tau \exp[-E_{\pi,A}(p)\tau] + (\tau \rightarrow N_t - \tau) 
  \\
  C_{{\rm meson},f_0}(\vec p,\tau) &=& c_{0}(p) + b_{f_0}(p)\exp[-E_{f_0}(p)\tau] + 
   b_{\eta,A}(p)(-)^\tau \exp[-E_{\eta,A}(p)\tau] + (\tau \rightarrow N_t - \tau). \nonumber
\end{eqnarray}
This fitting model adds explicit $a_0$ and $f_0$ poles, as well as the corresponding
negative parity states, to the bubble
contribution.  Such states are outside the scope of the low order
chiral Lagrangian in Eq.~(\ref{eq:SXPT}).  Of course it is possible to
enlarge the Lagrangian to include them \cite{Bardeen:2001jm}.
Taste-breaking effects complicate this exercise.  Moreover, we would
need to introduce a variety of higher order chiral couplings, which
are unlikely to be well constrained by our data.  Therefore, we took
the more modest approach and treated these additional terms
empirically, keeping in mind the possibility of higher order chiral
effects.

Our parameterization of the momentum dependence of the overlap factors
$b_j(p)$ requires some discussion.  The $a_0$ and $f_0$ are produced
through the scalar density with spin-taste assignment $1 \times 1$.
Thus at zeroth order in the $a_0 - \pi - \eta$ coupling their
contributions should be inversely proportional to their energies
$b_j(0) = 1/2E_j(p)$.  At higher order an iteration of the bubble
contribution alters the momentum dependence of the pole residue
\cite{Bardeen:2001jm}.  For present purposes we chose the empirical
fitting form
\begin{equation}
   b_j(p) =   b_{j0} + b_{j1}p^2.
\end{equation}
and adjusted the constants $b_{j0}$ and $b_{j1}$.

The negative parity states are the taste-axial-vector pion $\pi_A$ and
the taste-axial-vector $\eta_A$.  As staggered partners to the $a_0$
and $f_0$ they couple through axial vector currents with spin-taste
assignment $\gamma_0 \gamma_5 \times \gamma_0 \gamma_5$, which
contribute a factor of the energy to source and sink. Thus their bare
momentum dependence should be proportional to their energies
\begin{equation}
   b_j(p) =   b_{j}E(p).
\end{equation}
We kept this form, adjusting $b_j$.

The constant $c_{0}(p)$ is zero for all momenta except $\vec p =
(0,0,0)$, in which case it gives the vacuum-disconnected part of the
$f_0$ correlator.  There are eleven fit parameters for the meson terms
alone, but the two negative parity masses were constrained tightly by
priors: the $\pi_A$, to the previously measured value, and the $\eta_A$, to
the same derived mass that we used in the bubble term.

The bubble terms $B_{a_0}$ and $B_{f_0}$ in the fitting function Eq
(\ref{eq:fitfcn}) are given in momentum space by
Eqs.~(\ref{eq:a0bubblemom}) and (\ref{eq:f0bubblemom}).  Their
time-Fourier transforms yield $B_{a_0}(\vec p,\tau)$ and $B_{f_0}(\vec
p,\tau)$ by applying the following identity term by term:
\begin{equation}
   B(\vec p,\tau) \propto \frac{1}{A^2 (N_t/2)^2}\sum_{p_0,k}
    \frac{e^{-ip_0t}}{(k^2 + M_1^2)[(p - k)^2 + M_2^2]}
    =
    \sum_{\vec k}\frac{e^{-[E_1(\vec k) + E_2(\vec k)]t}}
   {4E_1(\vec k)E_2(\vec k)}
\end{equation}
where $E_j(\vec k) = \sqrt{|\vec k|^2 + M_j^2}$, and, as usual, $tA\equiv\tau a$.  Thus, for example, the
$\bar K K$ contribution to $B_{a_0}(\vec p,\tau)$ for taste $b$ is
\begin{equation}
  \frac{\mu^2}{16 L^3} \sum_{\vec k}
  \frac{e^{-[E_{Kb}(\vec k) + E_{Kb}(\vec k)]t}} 
  {4E_{Kb}(\vec k)E_{Kb}(\vec k)}.
\end{equation}

The bubble terms $B_{a_0}(p,\tau)$ and $B_{f_0}(p,\tau)$ were
parameterized by the three low energy couplings $\mu = m_\pi^2/(2
m_\ell)$, $\delta_A = a^4 \delta_A^\prime$, and $\delta_V = a^4
\delta_A^\prime$ in the notation of Ref.~\cite{Aubin:2004fs}.  They
were allowed to vary to give the best fit.  The taste multiplet masses
in the bubble terms were fixed as noted above.  The sum over
intermediate momenta was cut off when the total energy of the two-body
state exceeded $1.8/a$ or any momentum component exceeded $\pi/(3a)$.  We
determined that such a cut off gave acceptable accuracy for $\tau \ge
4$.

In summary, we fit all ten correlators with fourteen parameters,
eleven of which were needed to parameterize the four explicit meson
terms and three low energy couplings were needed for the bubble
contribution.  Through a prior, we constrained the value of $\delta_V$
to conform to previous fits to the pseudoscalar masses and decay
constants \cite{Aubin:2004fs}, leaving only two of the low energy
couplings to be adjusted independently.  Our best fit gave $\chi^2/dof
= 126/109$ (CL 0.13).

\begin{figure}[ht]
\epsfxsize=80mm
\begin{center}
\epsfbox{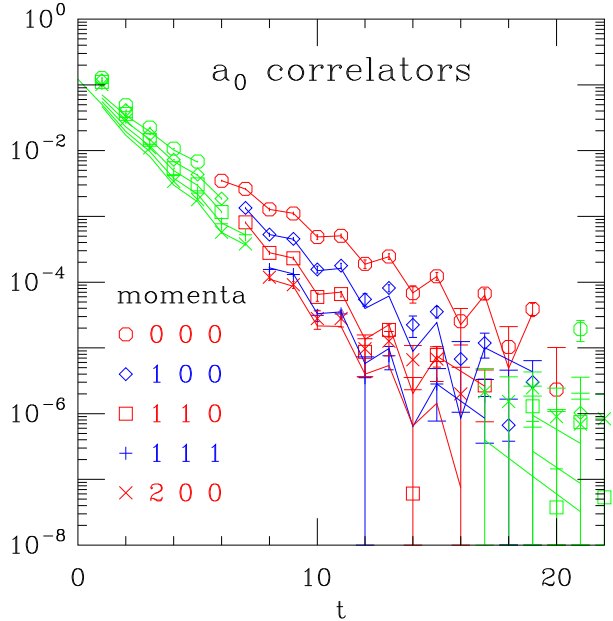}
\end{center}
\caption{Best fit to the $a_0$ correlator for five total cm
momenta. The fitting range is indicated by points and fitted lines in
red and blue (darker points and lines). Occasional points with
negative central values are not plotted.}  \label{fig:a0fit}
\end{figure}

\begin{figure}[ht]
\epsfxsize=80mm
\begin{center}
\epsfbox{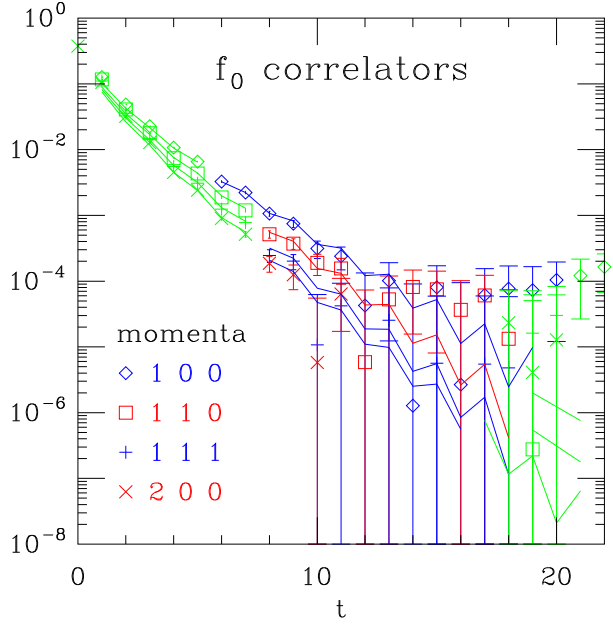}
\end{center}
\caption{Best fit to the $f_0$ correlator for four total cm momenta.}
\label{fig:f0fit}
\end{figure}

\begin{figure}[ht]
\epsfxsize=80mm
\begin{center}
\epsfbox{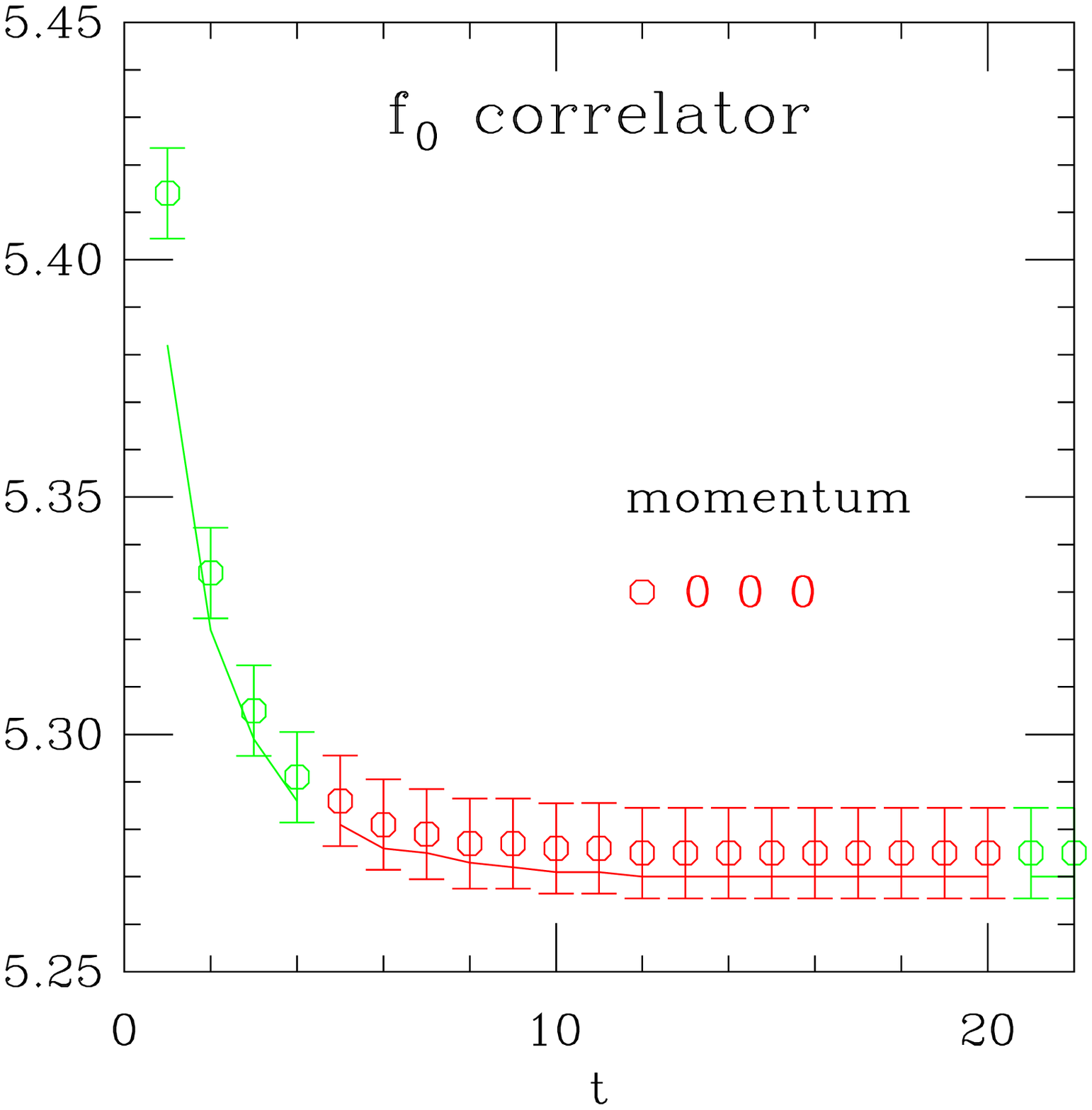}
\end{center}
\caption{Best fit to the zero momentum $f_0$ correlator.}
\label{fig:f0p0fit}
\end{figure}

The fitted functional form is compared with the data in
Figs~\ref{fig:a0fit}--\ref{fig:f0p0fit}.

Results of the fits are compared with results from fits to the meson
masses and decay constants in Table \ref{tab:LEC}.  The agreement is
worse if we used the bare value $r_1\mu = 4.5$ from those fits, rather
than the higher-order $m_\pi^2/(2 m_\ell)$, suggesting, perhaps, that
a higher order calculation of the bubble contribution might improve
the agreement.

The fitted masses of the $a_0$ and $f_0$ in units of the lattice
spacing are 0.61(5) and 0.45(9), respectively.

\begin{table}
  \begin{center}
  \begin{tabular}{lrr}
          & Our fit  & \qquad Meson masses and decays \\
  \hline
    $r_1 m_\pi^2/(2 m_{u,d})$ & 7.3(1.6) & 6.7 \\
    $\delta_V              $ & (prior)  & $-0.016(23)$ \\
    $\delta_A              $ & $-0.056(10)$ & $-0.040(6)$
  \end{tabular}
  \end{center}
  \caption{Comparison of our fit parameters for the \rSXPT\ low energy
    constants with results from \protect\cite{Aubin:2004fs} \label{tab:LEC}}
\end{table}


\section{Summary and Conclusions}
\label{sec:conclude}

We have derived the two-pseudoscalar-meson ``bubble'' contribution to
the $f_0$ correlator in lowest order \SXPT, thereby extending the
result for the $a_0$ in reference \cite{Prelovsek:2005rf}.  We have
used this model to fit simulation data for the point-to-point $a_0$
and $f_0$ correlators and found that best-fit values of the three
chiral low energy couplings are in reasonable agreement with values
previously obtained in fits to the light meson spectra and decay
constants \cite{Aubin:2004fs}.

The two-meson bubble term in \SXPT\ provides a useful illustration of
the lattice artifacts induced by the fourth-root approximation, since
it involves quark loops coming from the fermion determinant.  The
artifacts include thresholds at unphysical energies and thresholds
with negative weights.  These are the same sorts of artifacts commonly
observed with quenching or partial quenching.  These contributions are
clearly present in the $a_0$ and $f_0$ channels in our QCD simulation
with the Asqtad action at $a = 0.12$ fm.  We have found that they
must be taken into account in a successful spectral analysis.
Fortunately, \rSXPT\ provides an explicit parameterization of their
contributions for the interpolating operators we have chosen, thereby
allowing a fit to simulation data with a manageable number of
parameters.  The \rSXPT\ predicts further that these lattice artifacts
disappear in the continuum limit, leaving only physical two-body
thresholds.  This result is in full accordance with the fourth-root
analysis of Ref \cite{Bernard:2006zw}.  It will be interesting to see
whether this expectation is borne out in numerical QCD simulations at
smaller lattice spacing.


\section*{Acknowledgments}

This work is supported in part by the US National Science Foundation,
the US Department of Energy and Slovenian Ministry of Education,
Science and Sport.  We are grateful to the MILC Collaboration for the
use of the Asqtad lattice ensemble.  The analysis of these lattice
files was carried out at the Utah Center for High Performance
Computing.

\bibliographystyle{unsrt}
\bibliography{paper}

\begin{thebibliography}{10}

\bibitem{Bernard:2006ee}
Claude Bernard, Maarten Golterman, and Yigal Shamir.
\newblock Observations on staggered fermions at non-zero lattice spacing.
\newblock {\em Phys. Rev.}, D73:114511, 2006.

\bibitem{Bernard:2006zw}
C.~Bernard.
\newblock Staggered chiral perturbation theory and the fourth-root trick.
\newblock {\em Phys. Rev.}, D73:114503, 2006.

\bibitem{Shamir:2006nj}
Yigal Shamir.
\newblock Renormalization-group analysis of the validity of staggered-fermion
  qcd with the fourth-root recipe.
\newblock {\em Phys. Rev.}, D75:054503, 2007.

\bibitem{Sharpe:2006re}
Stephen~R. Sharpe.
\newblock Rooted staggered fermions: Good, bad or ugly?
\newblock {\em PoS}, LAT2006:022, 2006.

\bibitem{Bernard:2006qt}
Claude Bernard, Maarten Golterman, and Yigal Shamir.
\newblock Regularizing qcd with staggered fermions and the fourth root trick.
\newblock {\em PoS}, LAT2006:205, 2007.

\bibitem{Durr:2003xs}
Stephan Durr and Christian Hoelbling.
\newblock Staggered versus overlap fermions: A study in the schwinger model
  with n(f) = 0,1,2.
\newblock {\em Phys. Rev.}, D69:034503, 2004.

\bibitem{Follana:2004sz}
E.~Follana, A.~Hart, and C.~T.~H. Davies.
\newblock The index theorem and universality properties of the low- lying
  eigenvalues of improved staggered quarks.
\newblock {\em Phys. Rev. Lett.}, 93:241601, 2004.

\bibitem{Durr:2004as}
Stephan Durr, Christian Hoelbling, and Urs Wenger.
\newblock Staggered eigenvalue mimicry.
\newblock {\em Phys. Rev.}, D70:094502, 2004.

\bibitem{Durr:2004ta}
Stephan Durr and Christian Hoelbling.
\newblock Scaling tests with dynamical overlap and rooted staggered fermions.
\newblock {\em Phys. Rev.}, D71:054501, 2005.

\bibitem{Follana:2005km}
E.~Follana, A.~Hart, C.~T.~H. Davies, and Q.~Mason.
\newblock The low-lying dirac spectrum of staggered quarks.
\newblock {\em Phys. Rev.}, D72:054501, 2005.

\bibitem{Bernard:2005gf}
C.~Bernard et~al.
\newblock The locality of the fourth root of staggered fermion determinant in
  the interacting case.
\newblock {\em PoS}, LAT2005:114, 2006.

\bibitem{Bernard:2006wx}
C.~Bernard et~al.
\newblock Update on the physics of light pseudoscalar mesons.
\newblock {\em PoS}, LAT2006:163, 2007.

\bibitem{Aubin:2003mg}
C.~Aubin and C.~Bernard.
\newblock Pion and kaon masses in staggered chiral perturbation theory.
\newblock {\em Phys. Rev.}, D68:034014, 2003.

\bibitem{Aubin:2003uc}
C.~Aubin and C.~Bernard.
\newblock Pseudoscalar decay constants in staggered chiral perturbation theory.
\newblock {\em Phys. Rev.}, D68:074011, 2003.

\bibitem{Aubin:2004fs}
C.~Aubin et~al.
\newblock Light pseudoscalar decay constants, quark masses, and low energy
  constants from three-flavor lattice qcd.
\newblock {\em Phys. Rev.}, D70:114501, 2004.

\bibitem{Billeter:2004wx}
Brian Billeter, Carleton DeTar, and James Osborn.
\newblock Topological susceptibility in staggered fermion chiral perturbation
  theory.
\newblock {\em Phys. Rev.}, D70:077502, 2004.

\bibitem{Aubin:2004wf}
C.~Aubin et~al.
\newblock Light hadrons with improved staggered quarks: Approaching the
  continuum limit.
\newblock {\em Phys. Rev.}, D70:094505, 2004.

\bibitem{Gregory:2005yr}
Eric~B. Gregory, Alan~C. Irving, Craig~C. McNeile, Steven Miller, and Zbyszek
  Sroczynski.
\newblock Scalar glueball and meson spectroscopy in unquenched lattice qcd with
  improved staggered quarks.
\newblock {\em PoS}, LAT2005:027, 2006.

\bibitem{Prelovsek:2005qc}
Sasa Prelovsek.
\newblock Effects of partial quenching and staggered fermions on the scalar
  correlator.
\newblock {\em PoS}, LAT2005:085, 2006.

\bibitem{Prelovsek:2005rf}
S.~Prelovsek.
\newblock Effects of staggered fermions and mixed actions on the scalar
  correlator.
\newblock {\em Phys. Rev.}, D73:014506, 2006.

\bibitem{Bardeen:2001jm}
William~A. Bardeen, A.~Duncan, E.~Eichten, Nathan Isgur, and H.~Thacker.
\newblock Chiral loops and ghost states in the quenched scalar propagator.
\newblock {\em Phys. Rev.}, D65:014509, 2002.

\bibitem{Aubin:2003rg}
C.~Aubin and C.~Bernard.
\newblock Staggered chiral perturbation theory.
\newblock {\em Nucl. Phys. Proc. Suppl.}, 129:182--184, 2004.

\bibitem{Lee:1999zx}
Weon-Jong Lee and Stephen~R. Sharpe.
\newblock Partial flavor symmetry restoration for chiral staggered fermions.
\newblock {\em Phys. Rev.}, D60:114503, 1999.

\bibitem{Bernard:2001av}
Claude~W. Bernard et~al.
\newblock The qcd spectrum with three quark flavors.
\newblock {\em Phys. Rev.}, D64:054506, 2001.

\bibitem{Damgaard:2000gh}
P.~H. Damgaard and K.~Splittorff.
\newblock Partially quenched chiral perturbation theory and the replica method.
\newblock {\em Phys. Rev.}, D62:054509, 2000.

\bibitem{Dong:1993pk}
Shao-Jing Dong and Keh-Fei Liu.
\newblock Stochastic estimation with z(2) noise.
\newblock {\em Phys. Lett.}, B328:130--136, 1994.

\end{thebibliography}
\end{document}